\documentclass[aip,rsi,preprint,amsmath,amssymb,graphicx]{revtex4-1}

\let\ACMmaketitle=\maketitle
\renewcommand{\maketitle}{\begingroup\let\footnote=\thanks \ACMmaketitle\endgroup}

\usepackage{setspace}
\usepackage{multirow}
\usepackage{makecell}
\usepackage{rotating}
\usepackage{hyperref}
\hypersetup{colorlinks=true,citecolor=blue, linkcolor=blue, urlcolor=blue, breaklinks} 
\usepackage[left]{lineno}
\usepackage[T1]{fontenc}

\begin{document}


\title{A genetic algorithm approach to reconstructing spectral content from filtered x-ray diode array spectrometers\footnote{The following article has been submitted to by Review of Scientific Instruments. After it is published, it will be found \href{https://aip.scitation.org/journal/rsi}{here}.}
}  


\author{G.~E.~Kemp}
\email[]{kemp10@llnl.gov}
\affiliation{Lawrence Livermore National Laboratory, Livermore, CA 94550, USA}

\author{M.~S.~Rubery}
\affiliation{Atomic Weapons Establishment, Reading, RG7 4RS, UK}
\author{C.~D.~Harris}
\author{M.~J.~May}
\author{K.~Widmann}
\author{R.~F.~Heeter}
\author{S.~B.~Libby}
\author{M.~B.~Schneider}
\author{B.~E.~Blue}
\affiliation{Lawrence Livermore National Laboratory, Livermore, CA 94550, USA}


\date{\today}

\begin{abstract}

Filtered diode array spectrometers are routinely employed to infer the temporal evolution of spectral power from x-ray sources, but uniquely extracting spectral content from a finite set of broad, spectrally overlapping channel spectral sensitivities is decidedly nontrivial in these under-determined systems.  We present the use of genetic algorithms to reconstruct a probabilistic spectral intensity distribution and compare to the traditional approach most commonly found in literature. Unlike many of the previously published models, spectral reconstructions from this approach are neither limited by basis functional forms, nor do they require \emph{a priori} spectral knowledge.  While the original intent of such measurements was to diagnose the temporal evolution of spectral power from quasi-blackbody radiation sources -- where the exact details of spectral content was not thought to be crucial -- we demonstrate that this new technique can greatly enhance the utility of the diagnostic by providing more physical spectra and improved robustness to hardware configuration for even strongly non-Planckian distributions.

\end{abstract}


\maketitle 

\section{Introduction}
\label{sec:intro}

Filtered x-ray diode array spectrometers are routinely employed to characterize the x-ray emission \emph{vs.} time from high-energy density plasmas, such as those generated from laser-matter interactions\cite{1.1788872,1.1789603} or in Z-pinch devices\cite{PhysRevLett.81.4883}.  The spectral response, $R_i\left(E\right)$ (typically in units of $V/GW$) for photon energy $E$, of each diode channel $i$ is uniquely characterized by the choice of filter, diode, and (for low energy channels) x-ray mirror design.  By careful pairing of materials, channels can be strongly sensitive in (relatively) narrow spectral bands and an array of such channels can cover a broad spectral window from $eV$ to $keV$ photon energies.  The corresponding voltage trace recorded by channel $i$ is therefore described by

\begin{equation}
\label{eq:Vc}
V_i(t) = \Omega_i \int^{\infty}_{0} R_i\left(E\right) S\left(E,t\right) dE,
\end{equation}

\noindent where $\Omega_i$ is the detector solid angle and $S\left(E,t\right)$ is the incident spectral flux (typically in units of $GW/eV/sr$) at time $t$.  While conceptually straightforward to implement experimentally, uniquely recovering the incident spectral content from a finite set of channels is decidedly less trivial.  

Several spectral reconstruction algorithms have been suggested but most involve generating a finite set linear basis functions $\bold{B}$ such that $S\left(E\right) = \sum_{j}B_j\left(E\right)X_j$ where the weighting values $X_j$ are either solved for iteratively, through matrix inversion, or with non-linear least-square fitting \cite{doi:10.1063/1.2957935}.  Some of the more common forms of these basis functions include gaussians\cite{:/content/aip/journal/rsi/81/10/10.1063/1.3475385}, cubic-splines\cite{doi:10.1063/1.1785274}, or those generated through machine evolution\cite{cthomas}. While numerically fast and robust, the reconstructed spectral features are limited by the complexity and number of the basis functions and -- when compared to higher resolution spectral data\cite{doi:10.1063/1.4961267} -- can result in spurious spectral content. 

Other approaches suggest that more physical spectral reconstructions can be obtained when \emph{a priori} spectral information via candidate spectra -- provided either through modeling, experimental measurements, or with analytic physics models -- is incorporated into the reconstruction routine\cite{:/content/aip/journal/rsi/86/10/10.1063/1.4934542,doi:10.1063/1.4980151}.  The Levenberg-Marquardt method\cite{doi:10.1063/1.2957935}, for example, is routinely employed to analyze inertial confinement fusion hohlraums\cite{:/content/aip/journal/pop/11/2/10.1063/1.1578638} which models the x-ray emission as a linear combination of analytic bremsstrahlung, M-band (approximated as a gaussian bump), and Planckian (thermal) forms.  While the inclusion of \emph{a priori} information likely improves the fidelity of the extracted spectrum, the generalization of the algorithm to time-dependent or more complex x-ray sources becomes cumbersome. 

In this work, we introduce an alternative approach to extracting incident spectral content from filtered x-ray diode arrays using a Genetic Algorithm (GA).  Unlike many of the previously discussed routines, neither complex basis functional forms nor assumptions about the spectral content are explicitly necessary. In Section \ref{sec:ga}, we discuss the specifics of our implementation. Demonstration of the robustness and flexibility of the algorithm is provided through a variety of examples of physically representative x-ray sources using synthetic (\emph{i.e.} known) spectra and channel configurations in Section \ref{sec:synthetic}.  A comparison to the traditional approach most commonly employed in literature is also presented. We conclude with a discussion of the results in Section \ref{sec:discussion}.

\section{Genetic Algorithm Approach: \textsc{p\'uka}}
\label{sec:ga}

Genetic algorithms\cite{dejong1975} describe a suite of commonly employed heuristic techniques that have previously been demonstrated to be efficient at finding reasonable -- albeit not always optimal -- solutions to optimization problems where multiple local optima can exist and the parameter space is relatively large and unstructured\cite{Whitley1994}: \emph{e.g.} the traveling salesman problem\cite{Hoffman2013}. In essence, they mimic the evolutionary biological process of natural selection to iteratively produce a converged solution. The basic approach is as follows.  First, an initial population of potential solutions is generated -- each individual solution in the population has its own genome, an \emph{a priori} discretized representation of the solution domain.  Next, a goodness of fit parameter is used to sort and rank the genomes. A subsequent generation of solutions is then populated using a combination of routines; some fraction of the highest ranking genomes are selected to continue unaltered to next generation (known as elites) and the rest are altered from the previous generation, with the better solutions being given more opportunities to contribute genetic information to the next generation.  Some genomes are modified by combining two or more genomes (known as crossover) and others are mutations of a single parent genome -- multiple techniques exist for crossover and mutation\cite{Whitley1994}.  Each successive generation proceeds identically until the desired fit parameter or maximum number of allowed generations is reached.

Applying this technique to our filtered x-ray diode array spectrometer problem, the continuous incident spectral flux $S\left(E,t\right)$ becomes the genome.  We first discretize the spectrum and channel responsivities into a finite number $k$ of energy bins (\emph{i.e.} chromosomes), $S_k\left(E_k,t\right)$ and $R_{i,k}\left(E_k\right)$, respectively, to keep the problem numerically tractable.  The corresponding voltages are now calculated as 

\begin{equation}
\label{eq:Vd}
V_i(t) = \Omega_i \sum_{k} R_{i,k}\left(E_k\right) S_k\left(E_k,t\right) \Delta E_k,
\end{equation}

\noindent where $\Delta E_k$ are the bin widths whose spacing is chosen in an \emph{a priori} manner.  The population is initialized with mutations of the best-fitting Planckian distribution to the voltage traces where a variety of mutation routines (\emph{e.g.} single chromosome manipulation, chromosome pair flipping, add/subtract randomized gaussian and continuum spectral features, etc.) are employed.  The spectra are then ranked using a chi-squared fitness function given by

\begin{equation}
\label{eq:chi2}
\chi^2\left(t\right) = \frac{1}{N}\sum_{i=1}^{N} \frac{\left(V_i\left(t\right)- V_{i}^{m}\left(t\right)\right)^2}{V_i^{m}\left(t\right)},
\end{equation}

\noindent where $V_i^m$ is the measured voltage of channel $i$ for a system with $N$ channels.  For subsequent generations, elites typically constitute $1-2\%$ of the top ranked genomes, the top $50\%$ of the best fitting solutions are crossed through a variety of two-parent crossover routines (\emph{e.g.}, single and multiple point segment crossover, uniform mixing, weighted averaging, \emph{etc.}), and the rest are generated through mutation of both the elites and better fitting solutions; the remaining lower $50\%$ subsequently become extinct and do not contribute to the next generation.  Although the number of generations required for convergence will vary greatly depending upon the specifics of the GA implementation, we find convergence within roughly 500 generations for genomes with $300$ logarithmically-spaced energy bins and a population size of 100 for any of the spectra considered in this work. We note that this binning ansatz reasonably resolves all the relevant spectral features (\emph{e.g.} spectral edges and regions with large gradients) across all the channel response functions for the cases considered in this work.    

Due to the randomized nature of machine evolution and the infinite number of degenerate solutions that satisfy this under-determined -- yet consistent -- system of equations (\emph{i.e.} the number of energy bins exceeds the number of channels), the resultant spectra from separate reconstruction attempts never converge to a unique solution.  Indeed, GAs have previously been demonstrated to be capable of finding not just one, but many solutions to such systems of equations (when they exist)\cite{ikotun2011effectiveness}. In this scenario, probabilistic methods become necessary when reporting solutions\cite{jamisola2009probabilistic}. Therefore, multiple populations are generated and evolved in parallel. The resulting optimized solutions from each population are then consolidated to produce an averaged solution and standard deviation (both weighted by $1/\chi^2$). In this manner, only robust spectral features remain in the averaged result and uncertainties can be reported on a bin-by-bin basis. In this work, converged statistics from 100 different converged populations are presented as the probabilistic solution\footnote{We note that our implementation as a parallelized \textsc{python} script -- on a single node of one of Livermore's supercomputing machines with 36 cores/node -- takes about 100 seconds to run 100 separate reconstruction attempts.  The same calculation run in serial takes about 10 minutes.}. Due to the polymorphic nature of the routine, we have dubbed this algorithm \textsc{p\'uka} after the mischievous shape-shifting creature of Celtic folklore\cite{doi:10.1080/0015587X.1993.9715858}.

\section{Reconstructions of synthetic spectra}
\label{sec:synthetic}

Synthetic spectra from a variety of laser-matter interactions were generated to test the robustness of the genetic algorithm to representative, routinely diagnosed x-ray sources on both the OMEGA\cite{LPB:4326664} and National Ignition Facility (NIF)\cite{:/content/aip/journal/pop/11/2/10.1063/1.1578638} laser facilities. The simulations were performed in \textsc{hydra}\cite{Marinak:2001} -- a multi-physics, multi-dimensional, arbitrary Lagrangian-Eulerian, radiation-hydrodynamics code -- which incorporated non-local thermodynamic equilibrium (non-LTE) effects with detailed super-configuration accounting atomic models from \textsc{cretin}\cite{Scott201039}. Modeling details typical of such interactions can be found in prior publications\cite{:/content/aip/journal/pop/22/5/10.1063/1.4921250}. While physically meaningful spectra are not necessary to demonstrate the capabilities of the approach, we note that the filtered diode array spectrometers employed on such facilities are meticulously designed to diagnose similar spectral content. For the purposes of this study, we consider 3 test spectra of increasing spectral complexity: (i) non-Planckian continuum emission from a direct-drive exploding-pusher\cite{doi:10.1063/1.5025724}, (ii) non-LTE Xe L-shell ($\sim4-7\,keV$) source\cite{doi:10.1063/1.5015927}, and (iii) non-LTE Kr K-shell ($\sim13\,keV$) source\cite{doi:10.1063/1.5097960}.  

\textsc{Dante} is the primary filtered x-ray diode array spectrometer used on both NIF\cite{1.1788872} and OMEGA\cite{1.1789603} laser facility.  \textsc{Dante} has 18 ``XRD-31'' x-ray diodes\cite{pellinen2009response} with Al, Ni, and Cr photocathodes. Several channels employ grazing incidence x-ray mirrors to eliminate sensitivity to high-energy photons. For the purposes of this study, we consider 3 NIF channel configurations: the first from shot N170924-003, the second from N180129-001, and the third from N180604-002, as detailed in Table \ref{tab:dante1}, \ref{tab:dante2}, and \ref{tab:dante3}, respectively.  The first is typically employed to quantify Xe L-shell flux ($\sim4-7\,keV$), the second is traditionally fielded for diagnosing

\begin{table*}[!ht]
 \caption{\label{tab:dante1}\textsc{Dante-1} channel configuration from N170924-003.  The Ta grids are 25 $\mu m$ thick with 8.8\% open area.}
 \resizebox{\textwidth}{!}{\begin{tabular}{c c c c c c}
 \hline
 \hline
 Channel & Diode & Filters: & Mirror: & Solid angle & Peak sensitivity:\\
 number & material & material/thickness [$\mu m$] & material/angle [deg] &  [$\times10^{-6}\,sr$] & energy/width [eV]\\
 \hline
 1 & Al & Sc/0.8 + Ti/0.5 + Ta grid & SiO$_2$/3.5 & 0.5786 & 352.86/57.39 \\
 2 & Al & B/1.0 + CH/0.8 + Ta grid & aC/5 & 0.5929 & 167.65/26.63  \\
 3 & Al & Lexan/4.0 + Ta grid & SiO$_2$/3.5 & 0.5786 & 248.86/45.95 \\
 4 & Ni & V/2.0 & SiO$_2$/2.5 & 0.5603 & 476.28/48.01 \\
 5 & Cr & Co/1.4 + CH/0.2 + Ta grid & - & 0.6514 & 716.48/84.27\\
 6 & Cr & Cu/1.1 + Ta grid & - & 0.6514 & 828.97/136.24 \\
 7 & Ni & Mg/5.0 + Zn/1.0 + CH/0.1 + Ta grid& - & 0.6514 & 948.75/90.38  \\
 8 & Ni & Mg/22 + Ta grid & - & 0.6514 & 1184.11/157.27  \\
 9 & Ni & Al/20& - & 0.6514 & 1426.76/178.52 \\
 10 & Al & Si/20& - & 0.6514 & 1713.98/151.52  \\
 11 & Al & Ti/26& - & 0.6514 & 4187.27/837.45\\
 12 & Al & Saran/66& - & 0.6514 & 2593.02/308.65 \\
 13 & Al & Ag/4.4& - & 0.6514 & 2987.79/481.18 \\
 14 & Al & Lexan/2.0 + CaF$_2$/21.0 + CH/3.0& - & 0.6514 &3103.96/620.79 \\
 15 & Ni & Fe/30 & - & 0.6514 & 6186.51/1200.29  \\
 16 & Ni & Ti/26& - & 0.6514 & 4166.37/833.27 \\
 17 & Cr & Mn/26.4 + Ni/3.6 & - & 0.6514 & 6256.33/334.71 \\
 18 & Ni & Mn/26.4 + Ni/3.6 & - & 1.773 & 5852.86/910.28 \\
 \hline
 \hline
 \end{tabular}}
 \end{table*}
 
\begin{table*}[!h]
 \caption{\label{tab:dante2}\textsc{Dante-1} channel configuration from N180129-001.  The Ta grids are 25 $\mu m$ thick with 8.8\% open area.}
  \resizebox{\textwidth}{!}{\begin{tabular}{c c c c c c}
 \hline
 \hline
 Channel & Diode & Filters: & Mirror: & Solid angle & Peak sensitivity:\\
 number & material & material/thickness [$\mu m$] & material/angle [deg] &  [$\times10^{-6}\,sr$] & energy/width [eV]\\
 \hline
 1 & Al & Sc/0.8 + Ti/0.5& SiO$_2$/3.5 & 0.5740 & 349.85/60.32 \\
 2 & Al & B/1.5 + CH/1.2 & aC/5 & 0.5881 & 172.49/20.32 \\
 3 & Al & Lexan/4.0 + Ta grid & SiO$_2$/3.5 & 0.5740 & 249.55/45.33\\
 4 & Ni & V/1.75 & SiO$_2$/2.5 & 0.5558 & 472.28/53.0 \\
 5 & Cr & Co/1.4 + CH/0.2 + Ta grid & - & 0.6514 & 713.29/88.62\\
 6 & Cr & Cu/2.0+ Ta grid & - & 0.6514 & 863.62/93.2 \\
 7 & Ni & Mg/5.0 + Zn/0.65 + CH/0.1 + Ta grid& - & 0.6514 & 947.78/90.89  \\
 8 & Ni & Mg/22 + Ta grid & - & 0.6514 & 1181.55/160.04  \\
 9 & Ni & Al/20 + Ta grid & - & 0.6514 & 1427.6/177.44 \\
 10 & Al & Si/20& - & 0.6514 & 1721.08/148.31  \\
 11 & Al & Fe/1.0 + Cr/0.65 + Parylene/5.0 + Ta grid& - & 0.6514 & 3547.61/709.52 \\
 12 & Al & Saran/66& - & 0.6514 & 2577.59/328.57 \\
 13 & Al & Ag/4.4& - & 0.6514 & 3002.96/462.74 \\
 14 & Al & Ti/13& - & 0.6514 & 3785.2/757.04 \\
 15 & Ni & Zn/25 & - & 0.6514 & 8998.97/750.87  \\
 16 & Ni & Al/250& - & 0.6514 & 14400/2874.49 \\
 17 & Cr &Ni/10 & - & 0.6514 & 7173.62/1355.65 \\
 18 & Ni & Fe/1.0 + Cr/0.65 + Parylene/5.0 & - & 1.773 &3337.92/667.58 \\
 \hline
 \hline
 \end{tabular}}
 \end{table*}
   
  \begin{table*}[!ht]
 \caption{\label{tab:dante3}\textsc{Dante-1} channel configuration from N180604-002.  The Ta grids are 25 $\mu m$ thick with 8.8\% open area.}
  \resizebox{\textwidth}{!}{\begin{tabular}{c c c c c c}
 \hline
 \hline
 Channel & Diode & Filters: & Mirror: & Solid angle & Peak sensitivity:\\
 number & material & material/thickness [$\mu m$] & material/angle [deg] &  [$\times10^{-6}\,sr$] & energy/width [eV]\\
 \hline
 1 & Al & Sc/0.8 + Ti/0.5 + Ta grid& SiO$_2$/3.5 & 0.5740 & 353.32/57.22\\
 2 & Al & B/1.0 + CH/0.8 + Ta grid & aC/5 & 0.5929 & 166.7/27.72 \\
 3 & Al & Lexan/4.0 + Ta grid + Ta grid & SiO$_2$/3.5 & 0.5786 & 249.55/45.33\\
 4 & Ni & V/2.0 + Ta grid & SiO$_2$/2.5 & 0.5603 & 475.46/49.08 \\
 5 & Cr & Co/1.4 + CH/0.2 + Ta grid & - & 0.6514 & 711.48/90.88\\
 6 & Cr & Al/250 & - & 0.6514 & 14300/2854.72 \\
 7 & Ni & Mo/32& - & 0.6514 & 15400 3087.01  \\
 8 & Ni & Y/50 + Al/10 & - & 0.6514 & 13000/2598.06  \\
 9 & Ni & Ge/42 C/375& - & 0.6514 & 9746.42/1576.5 \\
 10 & Al & Si/20& - & 0.6514 & 1707.2/158.34  \\
 11 & Al & Mg/16.4 + Ni/3.6& - & 0.6514 & 5599.46/1119.89 \\
 12 & Al & Saran/66& - & 0.6514 & 2585.91/317.75\\
 13 & Al & Ag/4.4& - & 0.6514 & 2958.99/516.24 \\
 14 & Al & Mg/22 + Ta grid& - & 0.6514 & 1156.77/191.7\\
 15 & Ni & Al/20 & - & 0.6514 & 1427.94/176.79  \\
 16 & Ni & Mg/5.0 + Zn/1.0 + CH/0.1 + Ta grid& - & 0.6514 & 944.71/97.07 \\
 17 & Cr & Cu/1.1 + Ta grid & - & 0.6514 & 834.28/130.22 \\
 18 & Ni & Al/250 & - & 1.773 & 14400/2887.03 \\
 \hline
 \hline
 \end{tabular}}
 \end{table*}
 
 \begin{figure*}[!h]
\includegraphics[width=\textwidth]{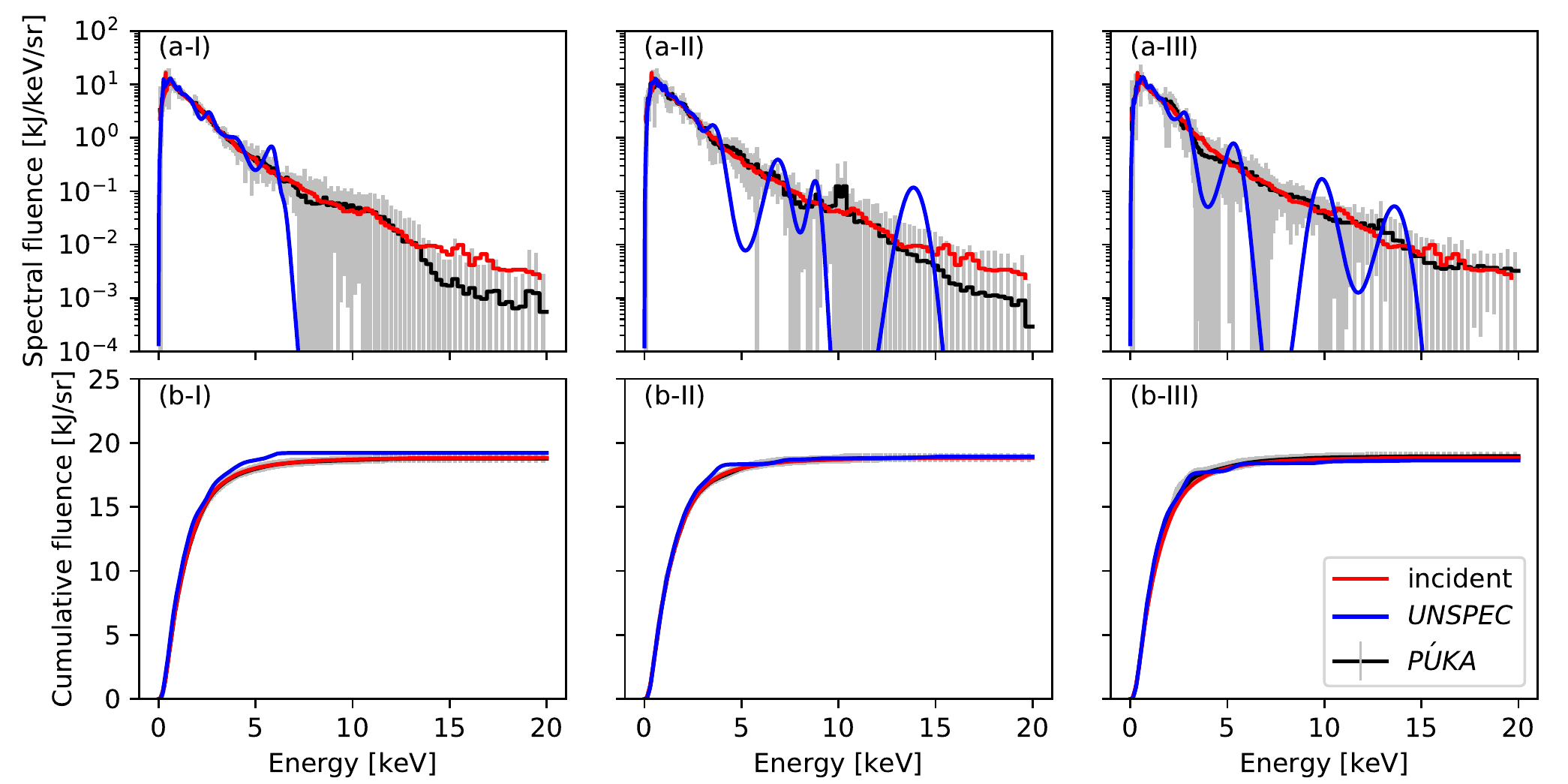}
\caption{\label{fig:N190227-001} Reconstructed (a) spectral and (b) cumulative spectral fluence for the non-Planckian continuum spectrum (i) using channel configurations \ref{tab:dante1}, \ref{tab:dante2}, and \ref{tab:dante3}.  Incident spectrum (red) is unknown to the reconstruction routines.}
\end{figure*}

\begin{figure*}[!h]
\includegraphics[width=\textwidth]{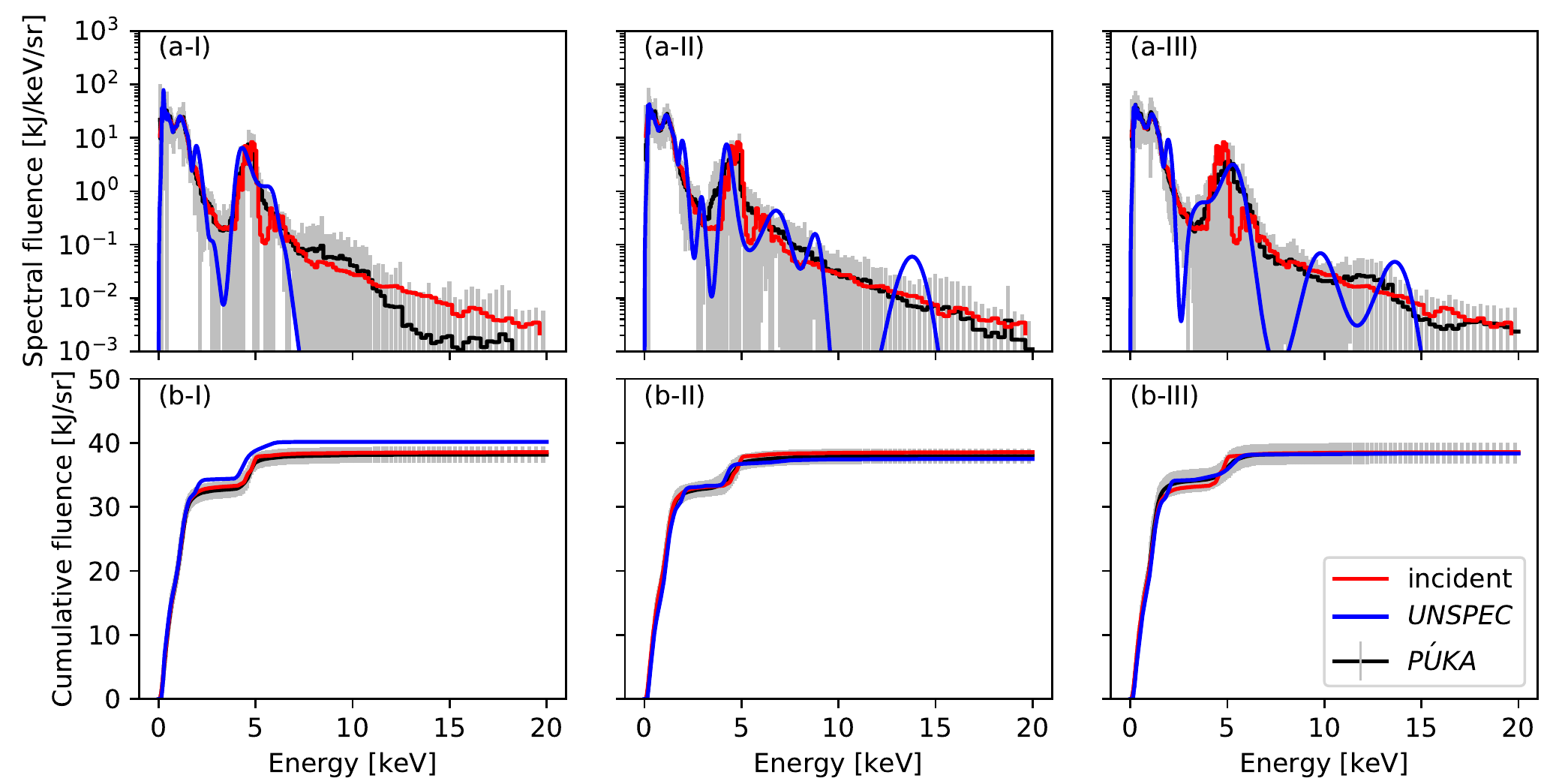} 
\caption{\label{fig:N170924-003} Reconstructed (a) spectral and (b) cumulative spectral fluence for the non-LTE Xe L-shell spectrum (ii) using channel configurations \ref{tab:dante1}, \ref{tab:dante2}, and \ref{tab:dante3}.  Incident spectrum (red) is unknown to the reconstruction routines.}
\end{figure*}

\begin{figure*}[!h]
\includegraphics[width=\textwidth]{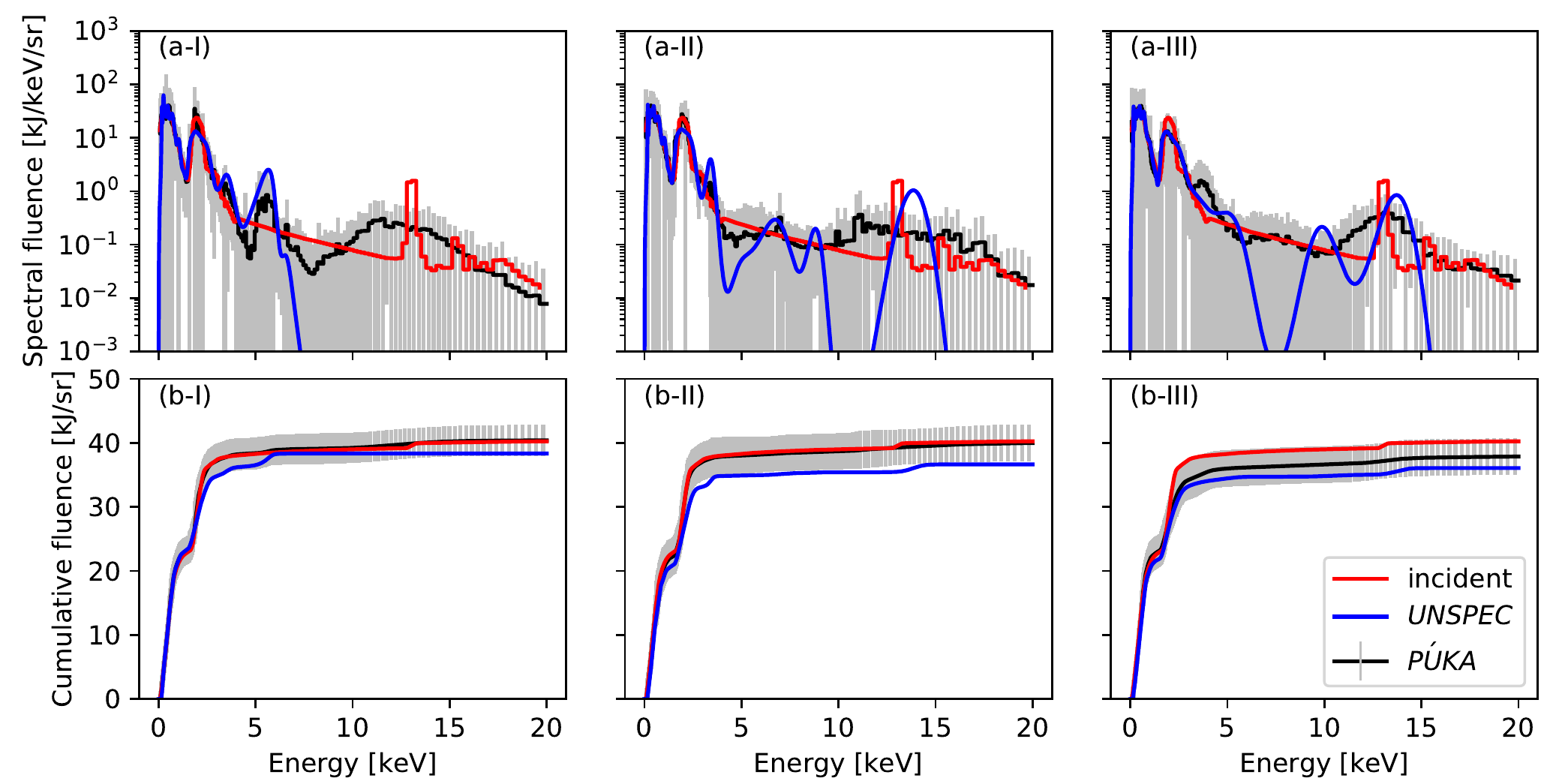} 
\caption{\label{fig:N141217-005} Reconstructed (a) spectral and (b) cumulative spectral fluence for the non-LTE Kr K-shell spectrum (iii) using channel configurations \ref{tab:dante1}, \ref{tab:dante2}, and \ref{tab:dante3}.  Incident spectrum (red) is unknown to the reconstruction routines.}
\end{figure*}
 
 \clearpage
  
\noindent hohlraum radiation temperature and M-shell flux ($\sim2-6\,keV$), and the third is usually employed to quantify Kr K-shell flux ($\sim13\,keV$).


Synthetic voltages were generated using (\ref{eq:Vd}) for each of the synthetic x-ray spectra and response functions described above. Spectral reconstructions from \textsc{p\'uka} and the traditional gaussian-basis method (known as \textsc{unspec}\cite{:/content/aip/journal/rsi/81/10/10.1063/1.3475385}\textsuperscript{,}\footnote{We note that the \textsc{unspec} results here used 100 iterations to ensure a converged solution, rather than the usual 10.}) are then compared and contrasted with the incident synthetic spectra (which is unknown to the algorithms); the results are shown in Figs. \ref{fig:N190227-001} and \ref{fig:N170924-003}, and \ref{fig:N141217-005}, respectively.  The uncertainty bars plotted with the \textsc{p\'uka} results indicate the $1/\chi^2$ weighted average $\pm2\sigma$ (standard deviation) for each energy bin for the (a) spectral fluence and (b) cumulative spectral fluence distributions. 

\section{Discussion}
\label{sec:discussion}

Outside of an initial Planckian guess, the traditional \textsc{unspec} routine only modifies spectral energy content via the fixed-position, fixed-width gaussian bumps in order to fit the observed voltages.  While this algorithmic ansatz has been demonstrated to work quite well for quasi-blackbody sources (at least in terms of the original intent of recovering total spectral power), it is often incapable of recovering sufficient spectral detail to recover the fluence metrics of interest for non-LTE spectra with bright bound-bound emission without meticulous channel design.  Consider, for example, the Kr K-shell spectrum shown in Fig. \ref{fig:N141217-005}; while \textsc{unspec} can recover the $>10\,keV$ K-shell fluence within 20\% (a typical value prescribed to the reconstructed fluences given traceable \textsc{Dante} calibration data\cite{:/content/aip/journal/rsi/81/10/10.1063/1.3475385}) with the specially designed channel configuration \ref{tab:dante3}, there are no gaussian bumps above $7\,keV$ in configuration \ref{tab:dante1} and, as such, little-to-no energy is attributed to the K-shell region -- an algorithmic fallacy to which \textsc{p\'uka} isn't subject. In general, we find that the reconstructed spectra from \textsc{p\'uka} are (i) more robust to channel selection and (ii) both qualitatively and quantitatively compare best with the incident spectral distribution and relevant broadband fluence metrics for each of the sources considered (summarized in Table \ref{tab:metrics}). Furthermore, the inherent probabilistic nature of \textsc{p\'uka} enables a dynamic and simultaneous assessment of the degree to which which spectral regions are constrained for a given configuration.

 \begin{table*}[!ht]
 \caption{\label{tab:metrics}Tabulated fluence metrics of typical interest for each type of x-ray source.  Uncertainty values for reconstructed \textsc{unspec} spectra are assumed to be $\pm20\%$. Reconstructed values in \textit{italic} font indicate an inability to recover the desired quantity within quoted uncertainty.}
  \resizebox{\textwidth}{!}{\begin{tabular}{c c c c c c}
 \hline
 \hline
 \multirow{2}{*}{Spectrum} & \multirow{2}{*}{Fluence Metric} 		& \multirow{2}{*}{Incident [kJ/sr]} 	& \multicolumn{3}{c}{Reconstructed [kJ/sr]} \\
                                          &                                                   		&                                                  	& Configuration & \textsc{p\'uka} & \textsc{unspec} \\
 \hline
 \multirow{3}{*}{(i) Non-Planckian continuum} 		& \multirow{3}{*}{Total ($0-20\,keV$)} 	& \multirow{3}{*}{18.8} 			& \multirow{3}{*}{\makecell{\ref{tab:dante1} \\ \ref{tab:dante2} \\ \ref{tab:dante3}}} 	& 18.8$\pm$0.3 & 19.2$\pm$3.8 \\
                              		&                                                          	& 			 	 			& 																& 18.9 $\pm$0.4 & 18.9$\pm$3.8 \\
  			     		& 							      	& 					 		& 																& 19.0$\pm$0.4 & 18.6$\pm$3.7 \\
\hline
\multirow{6}{*}{(ii) Non-LTE Xe} 		& \multirow{3}{*}{Total ($0-20\,keV$)} 	& \multirow{3}{*}{38.5} 			& \multirow{3}{*}{\makecell{\ref{tab:dante1} \\ \ref{tab:dante2} \\ \ref{tab:dante3}}} 	& 38.2$\pm$1.3 & 40.2$\pm$8.4 \\
                              		&                                                          	& 			 	 			& 																& 38.3$\pm$1.2 & 37.5$\pm$7.5 \\
  			     		& 							      	& 					 		& 																& 38.4$\pm$1.7 & 38.3$\pm$7.7 \\
					\cline{2-6}
					& \multirow{3}{*}{L-shell ($4-7\,keV$)} 	& \multirow{3}{*}{5.0} 			& \multirow{3}{*}{\makecell{\ref{tab:dante1} \\ \ref{tab:dante2} \\ \ref{tab:dante3}}} 	& 5.06$\pm$0.34 & 5.55$\pm$1.11 \\
                              		&                                                          	& 			 	 			& 																& 4.10$\pm$1.34 & \textit{3.63}$\pm$\textit{0.73} \\
  			     		& 							      	& 					 		& 																& \textit{3.87}$\pm$\textit{0.63} & \textit{3.55}$\pm${0.71} \\
\hline
\multirow{9}{*}{(iii) Non-LTE Kr} 		& \multirow{3}{*}{Total ($0-20\,keV$)} 	& \multirow{3}{*}{40.2} 			& \multirow{3}{*}{\makecell{\ref{tab:dante1} \\ \ref{tab:dante2} \\ \ref{tab:dante3}}} 	& 40.4$\pm$2.4 & 38.3$\pm$7.7 \\
                              		&                                                          	& 			 	 			& 																& 40.0$\pm$2.9 & 36.6$\pm$7.3 \\
  			     		& 							      	& 					 		& 																& 37.9$\pm$2.9 & 36.1$\pm$7.2 \\
					\cline{2-6}
					& \multirow{3}{*}{L-shell ($1.5-5\,keV$)} 	& \multirow{3}{*}{15.4} 			& \multirow{3}{*}{\makecell{\ref{tab:dante1} \\ \ref{tab:dante2} \\ \ref{tab:dante3}}}  	& 15.1$\pm$1.0 & 13.2$\pm$2.6 \\
                              		&                                                          	& 			 	 			& 																& 15.7$\pm$0.3 & 14.0$\pm$2.8 \\
  			     		& 							      	& 					 		& 																& \textit{12.8}$\pm$\textit{2.3} & \textit{12.6}$\pm$\textit{2.5} \\
					\cline{2-6}
					& \multirow{3}{*}{K-shell ($>10\,keV$)} 	& \multirow{3}{*}{1.2} 			& \multirow{3}{*}{\makecell{\ref{tab:dante1} \\ \ref{tab:dante2} \\ \ref{tab:dante3}}}  	& 1.17$\pm$0.13 & \textit{0.00}$\pm$\textit{0.00} \\
                              		&                                                          	& 			 	 			& 																& \textit{1.28}$\pm$\textit{0.06} & 1.23$\pm$0.24 \\
  			     		& 							      	& 					 		& 																& 1.26$\pm$0.08 & 1.21$\pm$0.24 \\
 \hline
 \hline
 \end{tabular}}
 \end{table*}

The \textsc{p\'uka} uncertainty values are indicative of the variability in the 100 degenerate solutions rather than any kind of experimental uncertainty. Individual bin uncertainties larger than the average indicate an inability of the algorithm to uniquely place energy -- a result of the reconstructed spectra having a higher local bin resolution than the system (\emph{i.e.} the combination of the diagnostic configuration and routine) is capable of robustly reconstructing.  Consider, for example, a region of the spectrum where the the response functions are largely parallel. It is not possible to uniquely determine the local distribution of energy in this scenario; the only way to better constrain the spectrum in this region is to modify the local channel responsivities (or provide additional constraints). While the resulting spectral uncertainties are inherently sensitive to choice of binning, the broadband fluence metrics of interest converge regardless of bin resolution (within reasonable limits). For example, running the same calculation with an order of magnitude less bins (30 vs. 300) does indeed significantly reduce the spectral uncertainty in some regions, but the cumulative spectral values and respective uncertainties are largely insensitive to this change. As such, we find that cumulative spectral fluence distributions can be a more robust and physically intuitive way of illustrating spectral uncertainty and overall energetic discrepancy with respect to the incident spectra. We note that these cumulative distributions are not generated from the averaged spectra and resulting standard deviations (\emph{i.e.} through a quadratic addition of errors) but rather the weighted average of the individual cumulative distributions themselves as the former cannot capture decreases in cumulative uncertainty.

Despite being more robust to channel configuration than the traditional \textsc{unspec} routine, the fidelity of the \textsc{p\'uka} reconstructions is still subject to the specifics of the spectral content. On occasion, \textsc{p\'uka} has been observed to robustly invent features in regions where spectral sensitivity is insufficient (\emph{e.g.} the feature around $\sim5\,keV$ in Fig. \ref{fig:N141217-005}(a)), albeit decidedly less spurious than some \textsc{unspec} reconstructions.  We note that robustness of spectral features doesn't necessarily imply physicality.  Without redesigning a tailored channel configuration, additional solution constraints (other than the only current requirement that spectral intensities be positive-definite) may become necessary to further suppress unphysical features.  In addition to applying \emph{a priori}/\emph{a posteriori} spectral knowledge, an adaptive binning routine or additional algorithmic constraints in low-sensitivity spectral regions may become necessary for some applications.

While due diligence is still necessary when designing an optimized experimental configuration for the spectral metrics of interest, we anticipate that this approach may alleviate some hardware driven design constraints originally imposed by the \textsc{unspec} routine.  As \textsc{p\'uka} doesn't necessitate narrow, spectrally independent channels, configurations with broadband energy responsivities that overlap significantly may improve the robustness of the configuration to the inevitability of experimental channel data loss  -- \emph{i.e.} when calibrated response functions are inconsistent with the resulting data (\emph{e.g.} broken filters) or when hardware issues arise (\emph{e.g.} inadequate scope settings or high-voltage bias on the diodes not being fielded as specified) -- as compared to \textsc{unspec}\cite{cdharris2020}. Despite the current narrowband \textsc{Dante} configuration design approach, we find that the current level of spectral responsivity overlap can often be sufficient for determining some inconsistent channel data when a single spectral distribution cannot simultaneously fit all the voltage data to within experimental uncertainty.

Although beyond the scope of this work, it is anticipated that high-resolution spectra may further improve the spectral reconstruction with this approach.  As we envision, the high-resolution spectral content can simply overwrite a portion of the genome and \textsc{p\'uka} will only be allowed to modify other spectral regions. High resolution spectra can be provided either as candidate spectra from atomic-kinetics models or from independent experimental measurements, such as the time-integrated \textsc{Virgil} crystal spectrometer ($1.5-6\,keV$) already implemented at NIF on \textsc{Dante-1}\cite{doi:10.1063/1.4961267}. In addition to comparing and contrasting to algorithms other than \textsc{unspec}, future work will also explore the robustness of the routine to experimental uncertainty: \emph{e.g.} those due to response function calibration uncertainty or channel mistiming in time-resolved reconstructions.  A Monte-Carlo approach would appear well suited to this task -- as has been previously implemented for \textsc{unspec}\cite{:/content/aip/journal/rsi/81/10/10.1063/1.3475385} -- but it is uncertain at this time how the degeneracy of the spectral solutions will also feed into this consideration. 


\section{Acknowledgments}
\noindent The first author would like to acknowledge many enlightening conversations with D.~Barnak, Y.~Frank, D.~A.~Liedahl, C.~W.~Mauche, E.~V.~Marley, A.~S.~Moore, P.~L.~Poole, and C.~A.~Thomas.  

\noindent This work performed under the auspices of U.S. Department of Energy by Lawrence Livermore National Laboratory under Contract DE-AC52-07NA27344. This document was prepared as an account of work sponsored by an agency of the United States government. Neither the United States government nor Lawrence Livermore National Security, LLC, nor any of their employees makes any warranty, expressed or implied, or assumes any legal liability or responsibility for the accuracy, completeness, or usefulness of any information, apparatus, product, or process disclosed, or represents that its use would not infringe privately owned rights. Reference herein to any specific commercial product, process, or service by trade name, trademark, manufacturer, or otherwise does not necessarily constitute or imply its endorsement, recommendation, or favoring by the United States government or Lawrence Livermore National Security, LLC. The views and opinions of authors expressed herein do not necessarily state or reflect those of the United States government or Lawrence Livermore National Security, LLC, and shall not be used for advertising or product endorsement purposes.

\section{Data availability}

\noindent The data that support the findings of this study are available from the corresponding author upon reasonable request.


%


\end{document}